# The Critical Magnetic Field of Anisotropic Two-Band Magnetic Superconductors

A.Changjan[1,2,3] and P.Udomsamuthirun[1,2]

[1] Prasarnmitr Physics Research Unit,Department of Physics, Faculty of Science ,Srinakharinwirot University Bangkok 10110,Thailand.
e-mail:udomsamut55@yahoo.com
[2] Thailand Center of Excellence in Physics, Si Ayutthaya Road,Bangkok 10400,Thailand.
[3] Department of Mathematics and Basic science, Faculty of Science and Technology ,Phathumwan Institute of Technology, Bangkok 10330,Thailand.

**PACS :74.20.Fg; 74.70.Ad**

## Abstract

The upper and lower critical field, and the critical field ratio of an anisotropic two-band magnetic superconductors in Ginzburg-Landau(GL) scenery is derived analytically. The temperature-dependent on the upper critical field is investigated and applied to Fe-based superconductors. We find that the very high value of zero-temperature upper critical field in Fe-based superconductors can be found in the negative differential susceptibility region. The temperature-dependent upper critical field is presented in two formulas as in the empirical view and in the GL two-band view that agrees with the experiment results.

**Keywords**: Upper critical field, Lower critical field, Ginzburg-Landau Theory, anisotropic magnetic superconductors, Two-Band superconductors

## 1.Introduction

The characteristic of superconductors in the magnetic field is widely studied in the framework of the Ginzburg-Landau(GL) theory [1,2,3]. For the non-magnetic superconductors, the magnetic response is diamagnetic that can be described well by GL theory. However, in the magnetic superconductors, the superconductivity can coexist with paramagnetic ion that shown the non-linear responds in the magnetic field. The properties of magnetic superconductors were considered by Hampshire[4,5] that the GL theory of one-band magnetic superconductors, including the spatial variation and nonlinear magnetic response of magnetic ions in-field was studied. He found that a ferromagnetism and antiferromagnetism can occur in the superconducting state. Askerzade[6] studied two-band GL theory and applied to determine the temperature-dependence lower, upper and thermodynamic critical field of non-magnetic superconductors, $LuNi_2B_2C$, $YNi_2B_2C$ and $MgB_2$. He showed the presence of two order parameters in the GL theory gave a non-linear temperature-dependence upper and lower critical magnetic field. Askerzade[7,8] , Udomsamuthirun et al.[9],Min-xia and Zi-Zhao[10] studied the upper critical field of anisotropic two-band superconductors by GL theory. The compact form of upper critical magnetic field of anisotropic two-band s-wave superconductor was found[9].

There are many kinds of the two-band superconductors found today. The two-band non-magnetic superconductors can be found in borocarbides material such as $LuNi_2B_2C$ ,$YNi_2B_2C$ and $MgB_2$. The two-band and multi-band magnetic superconductors that found in Fe-based superconductor was firstly found in $LaFeAsO_{1-x}F_x$ [11]. The material in this group has the oxypnictide form, $LnFeAsO$ here $Ln$ is rare earth elements. The anisotropic upper critical field of Fe-based superconductors is also found[12]. The very high upper critical field of the pnictide superconductors exceeds the capability of the common high field measurement system that the theoretical model is needed for helping the experimenter to predict this quantity. The experiments with powerful techniques such as ARPES, point-contact spectroscopy, and STM have been carried to study the superconducting gap in pnictides (for a review see[13]).The results are sometimes in disagreement with each other. For the point-contact spectroscopy measurements, the compound $LaFeAsO_{0.9}F_{0.1}$ with $T_c = 27$ K gave $\Delta_1(0) \approx 3.8$ meV and $\Delta_2(0) \approx 8.0$ meV[14] and the compound $SmFeAsO_{0.8}F_{0.2}$ with $T_c = 52$ K gave $\Delta_1(0) \approx 17.0$ meV and $\Delta_2(0) \approx 5.7$ meV [15,16]. For the ARPES measurement, the compound $Ba_{0.6}K_{0.4}Fe_2As_2$ with $T_c = 37$ K gave $\Delta_1(0) \approx 12.1$ meV and $\Delta_2(0) \approx 5.5$ meV and $\Delta_3(0) \approx 12.8$ meV [17]. In the scanning tunneling spectroscopy measurement, the compound $NdO_{0.86}F_{0.14}FeAs$ with $T_c = 48$ K gave $\Delta_1(0) \approx 18$ meV and $\Delta_2(0) \approx 9$ meV[18]. Ummarino[19] explained iron pnictides superconductor in the framework of s-wave three-band Eliashberg theory; two bands from holes band and one band from electron band. The s-wave order parameter of the hole bands have opposite sign with respect to that of the electron band ($s\pm$wave symmetry). Sadovskii[20] grouped the order parameter of iron pnictides superconductor and proposed the two main groups of order parameters that occurred in this materials, d-wave symmetry and $s\pm$wave symmetry. The d-wave symmetry is found in case those gaps on different pockets of electron-like sheets differ in sign, while two gaps on hole-like sheets are zero, and the $s\pm$wave symmetry occurred when gaps of electron-like pockets have the same signs, while the gaps of hole-like pockets have opposite signs. The possibility of $s\pm$wave symmetry in FeAs-based compounds was first proposed by Mazin et al.[21] that agreed with ARPES data[17]. The order parameters in this case were reduced into two bands. Day[22] suggested that the Fe-based superconductor could be d-wave symmetry or $s\pm$wave symmetry depending on the details of the material.

Koshelev and Golubov[23] showed that anisotropic GL theory does not apply well to two-band superconductor in $MgB_2$ because of the special shape of the Fermi surfaces. The superconductivity in $MgB_2$resides in two families of the band: strongly superconducting quasi-two-dimensional $\sigma$-bands and weakly superconducting three-dimensional $\pi$-bands and the symmetry of the superconducting state is s-wave symmetry. The criterion for the validity of the anisotropic GL theory is depended on the ratio of coherence length of $\sigma$-bands to of $\pi$-bands and the ratio of parameters that related to the coupling constants. In $MgB_2$, these parameters give the validity of the anisotropy GL theory limited to extremely narrow temperature range near $T_c$. In pnictides superconductor, there are theories[19-21] suggested that they are $s\pm$wave paring, in which the gaps on different Fermi surface sheets have different signs. The Fermi surface consists of five sheets: two electron cylinders, two hole cylinders and heavy three-dimensional hole pocket. The electron cylinders make the dominant contribution to the in-plane electrical conductivity. The superconductivity parameter for cuprate and pnictides are in the same size class[24].Their phase diagrams consist of superconducting phase and magnetic phase, in contrast to $MgB_2$. The magnetic excitations lead to paring mechanism that in $MgB_2$ is the electron-phonon interaction. However in many respects, pnictides sit between cuprate and $MgB_2$ superconductors[25]. We can assume that the ratio of the coherence and coupling constant of pnictides are in same class of cuprate that the region of applicability of the anisotropy GL theory exists. The anisotropic GL can apply to pnictides.

In this paper, we calculate analytically the upper and lower critical field, and the critical field ratio of an anisotropic magnetic superconductor by Ginzburg-Landau theory. The temperature-dependent on upper critical field are also investigate and applied to Fe-based superconductors in case of $s\pm$wave symmetry.

## 2.Model and Calculations

The Helmholtz free energy of Ginzburg and Landau[1] theory can be modified to describe the critical magnetic field in non-magnetic superconductors and magnetic superconductors both in one-band and two-band presented in Refs.[1-10,26-28]. For isotropic two-band GL theory, there are two order parameters in the Ginzburg-Landau functional that free energy can be written down as

$$F_{sc}[\psi_1,\psi_2]=\int d^3r f_{sc}=\int d^3r\left(F_1+F_2+F_{12}+\int H_s dB\right) \tag{1}$$

with
$$F_{i(i=1,2)}=\frac{1}{2m_i}\left|\left(-i\hbar\vec{\nabla}-2e\vec{A}\right)\psi_i\right|^2+\alpha_i|\psi_i|^2+\frac{1}{2}\beta_i|\psi_i|^4 \tag{2}$$

$$F_{12}=\varepsilon\left(\psi_1^*\psi_2+c.c.\right)+\varepsilon_1\left\{\left(i\hbar\vec{\nabla}-2e\vec{A}\right)\psi_1^*\left(-i\hbar\vec{\nabla}-2e\vec{A}\right)\psi_2+c.c.\right\} \tag{3}$$

Here $F_i$ is the free energy of the separate bands. $F_{12}$ is the interaction term between bands. $m_i$ denotes the effective mass of carriers. $\psi_i$ is the order parameter and $|\psi_i|^2$ is proportional to the density of carriers, the coefficient $\alpha_i$ depends on the temperature, while coefficient $\beta_i$ is independent of temperature. The first term in Eq.(1) accounts for the kinetic energy of the carriers and the lasted term accounts for the energy stored in the local magnetic fields. $H_s=\frac{B}{\mu_0}-M_{ions}$, $M=M_{sc}+M_{ions}$ and $M_{ions}=\chi H_s$, where $\mu_0 M_{sc}$ is the field produced by the carriers and $\mu_0 M_{ions}$ is the field produced by the ions. When the magnetic superconductors place in magnetic field, there are the field produced by ions $(\mu_0 M_{ions})$, the applied field $(\mu_0 H)$ and the field produced by the carriers $(\mu_0 M_{sc})$ that can produce relationship between field as $B=\mu_0 H+\mu_0 M_{sc}+\mu_0 M_{ions}$ and $M_{ions}=\chi H_{c_2}(T)+\chi'\left(H-M_{sc}-H_{c_2}\right)$. So that $B=\mu_0(\chi-\chi')H_{c_2}+\mu_0(1+\chi')(H+M_{sc})$ where the differential susceptibility $(\chi')$ is $\left(\frac{\partial M_{ions}}{\partial H}\right)_{T,H=H_{c_2}}$ and susceptibility $(\chi)$ is $\left(\frac{M_{ions}}{H}\right)_{T,H=H_{c_2}}$.

Therefore, the free energy of an isotropic two-band magnetic superconductor can be written as

$$F_{sc}[\psi_1,\psi_2]=\int d^3r\left(F_1+F_2+F_{12}+\int\left(B-\mu_0 M_{ions}\right)\frac{dB}{\mu_0}-\left(B-\mu_0 M\right)M\right) \tag{4}$$

For small change in B-field, a series in B is introduced [4, 5, 26]

$$\gamma_0+\gamma_1 B+\gamma_2\frac{B^2}{2\mu_0}=\int\left(B-\mu_0 M_{ions}\right)\frac{dB}{\mu_0}-\left(B-\mu_0 M\right)M_{sc}-\left(B-\mu_0 M\right)M_{ions} \tag{5}$$

Here $\gamma_0,\gamma_1$ and $\gamma_2$ are coefficient parameters and $\gamma_2=1$ for non-magnetic superconductors. The coefficient parameter $\gamma_1$ depends on the gradient of field and $\gamma_1=0$ for the uniform applied field.

The anisotropic gap function can be written as [29] $\Delta(\hat{k},T) \sim f(\hat{k})\Delta(T)$ where $f(\hat{k})$ is the anisotropic function. Similarity, we can write anisotropic order parameter as

$$\psi(\hat{k},T) = f(\hat{k})\psi(T) \tag{6}$$

There are two anisotropic gap functions with uniaxial symmetry proposed by Haas and Maki[30]: $f(\theta') = \dfrac{1 + a\cos^2\theta'}{1+a}$ and proposed by Posazhennikova et al.[31]: $f(\theta') = \dfrac{1}{\sqrt{1 + a\cos^2\theta'}}$ ,here $\theta'$ is the polar angle and $a$ is anisotropic parameter. In case of the symmetry order parameter, $f(\hat{k}) = 1$.

The free energy of anisotropic two-band magnetic superconductors can be obtained as

$$F_{sc}[\psi_1,\psi_2] = \int d^3r f_{sc} = \int d^3r\left(F_1 + F_2 + F_{12} + \gamma_0 + \gamma_1 B + \gamma_2\frac{B^2}{2\mu_0}\right) \tag{7}$$

with

$$F_{i(i=1,2)} = \frac{1}{2m_i}\left|\left(-i\hbar\vec{\nabla} - 2e\vec{A}\right)\psi_i\right|^2\left\langle f_i^2(\hat{k})\right\rangle + \alpha_i|\psi_i|^2\left\langle f_i^2(\hat{k})\right\rangle + \frac{1}{2}\beta_i|\psi_i|^4\left\langle f_i^4(\hat{k})\right\rangle \tag{8}$$

$$F_{12} = \varepsilon\left(\psi_1^*\psi_2 + c.c.\right)\left\langle f_1(\hat{k})f_2(\hat{k})\right\rangle + \varepsilon_1\left\{\left(i\hbar\vec{\nabla} - 2e\vec{A}\right)\psi_1^*\left(-i\hbar\vec{\nabla} - 2e\vec{A}\right)\psi_2 + c.c.\right\}\left\langle f_1(\hat{k})f_2(\hat{k})\right\rangle \tag{9}$$

The Eq.(7) is the general function that can describe all cases mentioned before.

Minimising $F_{sc}$ of Eq.(7) with respect to $\psi_i^*$ ($i = 1,2$) and $\vec{A}$, the 1st and 2nd GL equations including the anisotropic function and magnetic ions can be found as :

$$\begin{aligned}&\frac{\left\langle f_1^2(\hat{k})\right\rangle}{2m_1}\left(-i\hbar\vec{\nabla} - 2e\vec{A}\right)^2\psi_1 + \alpha_1\left\langle f_1^2(\hat{k})\right\rangle\psi_1 + \beta_1\left\langle f_1^4(\hat{k})\right\rangle\left(\psi_1^*\psi_1\right)\psi_1 \\ &\quad + \varepsilon\left\langle f_1(\hat{k})f_2(\hat{k})\right\rangle\psi_2 - \varepsilon_1\left\langle f_1(\hat{k})f_2(\hat{k})\right\rangle\left(-i\hbar\vec{\nabla} - 2e\vec{A}\right)^2\psi_2 = 0\end{aligned} \tag{10}$$

$$\begin{aligned}&\frac{\left\langle f_2^2(\hat{k})\right\rangle}{2m_2}\left(-i\hbar\vec{\nabla} - 2e\vec{A}\right)^2\psi_2 + \alpha_2\left\langle f_2^2(\hat{k})\right\rangle\psi_2 + \beta_2\left\langle f_2^4(\hat{k})\right\rangle\left(\psi_2^*\psi_2\right)\psi_2 \\ &\quad + \varepsilon\left\langle f_1(\hat{k})f_2(\hat{k})\right\rangle\psi_1 - \varepsilon_1\left\langle f_1(\hat{k})f_2(\hat{k})\right\rangle\left(-i\hbar\vec{\nabla} - 2e\vec{A}\right)^2\psi_1 = 0\end{aligned} \tag{11}$$

And

$$-\left(\frac{\gamma_1}{B}+\frac{\gamma_2}{\mu_0}\right)\mu_0\vec{\nabla}\times(1+\chi')M_{sc}=\frac{\left\langle f_1^2(\hat{k})\right\rangle}{2m_1}\left[2ie\hbar\left(\psi_1^*\vec{\nabla}\psi_1-\psi_1\vec{\nabla}\psi_1^*\right)+8e^2\vec{A}|\psi_1|^2\right]$$
$$+\frac{\left\langle f_2^2(\hat{k})\right\rangle}{2m_2}\left[2ie\hbar\left(\psi_2^*\vec{\nabla}\psi_2-\psi_2\vec{\nabla}\psi_2^*\right)+8e^2\vec{A}|\psi_2|^2\right] \quad (12)$$
$$+\varepsilon_1\left\langle f_1(\hat{k})f_2(\hat{k})\right\rangle\begin{bmatrix}2ie\hbar\left(\psi_1^*\vec{\nabla}\psi_2-\psi_2\vec{\nabla}\psi_1^*\right)+8e^2\vec{A}\psi_1^*\psi_2\\+2ie\hbar\left(\psi_2^*\vec{\nabla}\psi_1-\psi_1\vec{\nabla}\psi_2^*\right)+8e^2\vec{A}\psi_2^*\psi_1\end{bmatrix}$$

Where $\vec{\nabla}\times M_{sc}=J_{sc}$ is the supercurrent density and $<...>$ is averaged over Fermi surface.

### 2.1 Upper critical magnetic field($B_{c2}$)

Let there has the uniform applied magnetic field in z-direction, the vector potential can be set in the form of

$$\vec{A}=\left(0,\left[\mu_0(\chi-\chi')H_{c2}+\mu_0(1+\chi')(H+M_{sc})\right]x,0\right).$$

For convenient of calculation, the mass of carriers in each band $m_i$ ($i=1,2$) should be $m_1=m_2=m$. Because of the uniform applied field, the upper critical field can be obtained approximately from the linearised 1st GL equation, Eq.(10) that

$$B_{c2}=\mu_0H_{c2}=-\frac{m}{\hbar e(1+\chi)}\left\{\frac{\left[\alpha_1+\alpha_2+4m\varepsilon\varepsilon_1\Omega\right]}{\left(1-4m^2\varepsilon_1^2\Omega\right)}+\frac{\left(\alpha_1\alpha_2-\Omega\varepsilon^2\right)}{\left[\alpha_1+\alpha_2+4m\varepsilon\varepsilon_1\Omega\right]}\right\} \quad (13)$$

Here $B_{c2}$ is the upper critical magnetic field, and $\Omega=\frac{\left\langle f_1(\hat{k})f_2(\hat{k})\right\rangle^2}{\left\langle f_1^2(\hat{k})\right\rangle\left\langle f_2^2(\hat{k})\right\rangle}$ is a dimensionless parameter of anisotropy obtained in Ref.[9]and $\Omega$ is equal to 1 for the isotropic case.

The Eq.(13) shows that the presence of the magnetic ions reduces the upper critical field strength by a factor $(1+\chi)$ and the upper critical field is depended on the anisotropic parameter.

### 2.2 Lower critical magnetic field ($B_{c1}$)

Consider the uniform applied magnetic field just above $B_{c_1}$, there have the flux penetrations the superconductors. For the lower critical field's calculation that keeping as much symmetry as possible, we choose a flux line in cylindrical coordinate $\vec{B}=B(r)\hat{z}$ where $B(r)$ has its maximum value at a core and tends to zero at large radial. A vector potential in cylindrical coordinate is chosen in the same manner of $B_{c2}$'s calculation as

$$\vec{A}=\left[\mu_0(\chi-\chi')H_{c2}r+\mu_0(1+\chi')(H+M_{sc})r\right]\hat{\theta}=A(r)\hat{\theta} \quad (14)$$

The solution of the order parameters in this form $\psi_1(T)$ and $\psi_2(T)$ are provided by

$$\psi_1(T)=|\psi_1(T)|e^{i\phi_1} \quad \text{and} \quad \psi_2(T)=|\psi_2(T)|e^{i\phi_2}$$

Where $\phi_1$ and $\phi_2$ are the phases of the order parameters.

By Maxwell's equation, $\vec{j}=\frac{1}{\mu_0}\vec{\nabla}\times\vec{B}$ for the magnetostatic case; $\frac{\partial\vec{D}}{\partial t}=0$ and 2nd GL equation in uniform field, the vector potential is

$$\vec{A}=-\frac{w_1\vec{\nabla}\phi_1+w_2\vec{\nabla}\phi_2+w_3\left(\vec{\nabla}\phi_1+\vec{\nabla}\phi_2\right)}{w_4} \tag{15}$$

Here $\Phi=\frac{2\pi}{w_4}$ is the magnetic flux quantum that

$w_4=-\frac{2e}{\hbar}\left(w_1+w_2+2w_3\right)$ and

$$w_1=-\frac{\left\langle f_1^2(\hat{k})\right\rangle}{2m_1}4e\hbar\left(-\frac{\left(\alpha_1\left\langle f_1^2(\hat{k})\right\rangle+\frac{\varepsilon\left\langle f_1(\hat{k})f_2(\hat{k})\right\rangle}{\theta}\right)}{\beta_1\left\langle f_1^4(\hat{k})\right\rangle}\right),$$

$$w_2=-\frac{\left\langle f_2^2(\hat{k})\right\rangle}{2m_2}4e\hbar\left(-\frac{\left(\alpha_2\left\langle f_2^2(\hat{k})\right\rangle+\varepsilon\left\langle f_1(\hat{k})f_2(\hat{k})\right\rangle\theta\right)}{\beta_2\left\langle f_2^4(\hat{k})\right\rangle}\right),$$

$$w_3=-\varepsilon_1\left\langle f_1(\hat{k})f_2(\hat{k})\right\rangle 4e\hbar\left(\frac{\alpha_1\left\langle f_1^2(\hat{k})\right\rangle+\frac{\varepsilon\left\langle f_1(\hat{k})f_2(\hat{k})\right\rangle}{\theta}}{\beta_1\left\langle f_1^4(\hat{k})\right\rangle}\right)^{\frac{1}{2}}\left(\frac{\alpha_2\left\langle f_2^2(\hat{k})\right\rangle+\varepsilon\left\langle f_1(\hat{k})f_2(\hat{k})\right\rangle\theta}{\beta_2\left\langle f_2^4(\hat{k})\right\rangle}\right)^{\frac{1}{2}},$$

here $\theta=\frac{\psi_1}{\psi_2}$.

The equation of vector potential takes the form $\vec{j}=\vec{\nabla}\times\vec{b}$ , here $\vec{b}$ is internal field. We can get

$$\vec{\nabla}\times\vec{\nabla}\times\vec{b}=\frac{1}{-\gamma_2(1+\chi')}\left[\frac{\left\langle f_1^2(\hat{k})\right\rangle}{2m_1}8e^2|\psi_1|^2+\frac{\left\langle f_2^2(\hat{k})\right\rangle}{2m_2}8e^2|\psi_2|^2+\varepsilon_1\left\langle f_1(\hat{k})f_2(\hat{k})\right\rangle 8e^2\left(\psi_1^*\psi_2+\psi_2^*\psi_1\right)\right] \tag{16}$$

Let our sample has a surface perpendicular to the x-axis and the external magnetic field is $\vec{B}=B_0\hat{z}$, and the internal magnetic field can be the form $\vec{b} = b(x)\hat{z}$. Then, the London equation is of the form $\frac{d^2 b(x)}{dx^2} - \frac{1}{\lambda^2} b(x) = 0$. Here $\lambda$ is the anisotropic magnetic London penetration depth of the following form

$$\lambda^{-2} = \frac{1}{\gamma_2(1+\chi')}\left[\frac{2e}{\hbar}w_1 + \frac{2e}{\hbar}w_2 + \frac{4e}{\hbar}w_3\right] \quad . \tag{17}$$

The lower critical field can be obtained as

$$B_{c1} = \frac{\gamma_2(1+\chi')\hbar}{8e\mu_0\lambda^2}\ln\kappa \quad . \tag{18}$$

Here $\kappa$ is Ginzburg-Landau parameter; $\kappa = \frac{\lambda}{\xi}$, $\xi$ is coherence length that

$$\xi^2 = \frac{\langle f_1^2(\hat{k})\rangle\hbar^2}{2m_1|\alpha_1|} - E\frac{\langle f_2^2(\hat{k})\rangle\hbar^2}{2m_2|\alpha_2|} \quad \text{with} \quad E = -\frac{|\alpha_1|}{|\alpha_2|} \ .$$

We find that our lower critical field, Eq.(18), has the similar form to Abrikosov[2] but there is the difference in London penetration depth and coherence length and our result shows the anisotropic and magnetic parameters' dependence.

## 2.3 The critical magnetic field ratio ($\eta$)

We have the critical magnetic field ratio [26] that show the relation of $B_{c1}$ and $B_{c2}$ as

$$\eta = \frac{B_{c2}}{B_{c1}} \tag{19}$$

This ratio will help experimenter to predict the unknown parameter such as $B_{c2}$ of Fe-based superconductors that having very high $B_{c2}$.

Substitution Eq.(13) and Eq.(18) in Eq.(19), the critical magnetic field ratio is found as

$$\eta = \left[\frac{\mu_0}{(1+\chi)} \cdot \frac{\ln\kappa_0}{\ln\Theta\kappa_0}\right]\eta_0 \tag{20}$$

Where $\eta_0$ is the anisotropic non-magnetic superconductors that $\gamma_2 = \mu_0 = 1$, $\chi = \chi' = 0$, and

$$\eta_0 = \frac{8\lambda_0^2 m}{\hbar^2 \ln\kappa_0}\left[\frac{\alpha_1 + \alpha_2 + 4m\varepsilon\varepsilon_1\Omega}{1 - 4m^2\varepsilon_1^2\Omega} + \frac{\alpha_1\alpha_2 - \Omega\varepsilon^2}{\alpha_1 + \alpha_2 + 4m\varepsilon\varepsilon_1\Omega}\right] \tag{21}$$

Here $\Theta^2 = \gamma_2(1+\chi')$ is the magnetic parameter. $\kappa_0$ is the anisotropic non-magnetic GL parameter; $\kappa_0 = \frac{\lambda_0}{\xi_0}$ that $\lambda_0, \xi_0$ is the penetration depth and coherence length of an anisotropic non-magnetic superconductors, respectively.

We can get the simplify upper critical field equation as

$$B_{c2} = \left[\frac{\mu_0}{(1+\chi)} \cdot \frac{\ln \kappa_0}{\ln \Theta \kappa_0}\right] \eta_0 \cdot B_{c1} \qquad (22)$$

## 2.4 The temperature-dependence upper critical field

The temperature-dependence upper critical field, $B_{c2}(T)$, is one of the most interesting property in this field and there are a lot of experimental data on this. We can investigate the temperature-dependence of Eq.(22) by considering on the $\alpha$ and $\varepsilon$ that are the temperature-dependence parameters. In the GL theory, we have $\xi^2 \equiv \frac{1}{\alpha}$ and $\xi^2 \equiv \frac{1}{1-t}$ ,here $\xi$ is the coherence length and $t = \frac{T}{T_c}$, for near critical temperature region. So we cannot use this relation to estimate the upper critical field at zero-temperature, $B_{c2}(0)$. The $B_{c2}(T)$ that can extend to zero-temperature was proposed by Zhu et al.[32] that $\xi^2 \equiv \frac{1+t^2}{1-t^2}$ empirically. In our two-band GL, we make the assumption that our coherence length has the temperature-dependence of this form. As in Ref.[9], we have $\xi^2 \equiv \frac{1}{\alpha}$ and $\xi^2 \equiv \frac{1}{\varepsilon}$ ,then we get $\alpha \equiv \frac{1-t^2}{1+t^2}$ and $\varepsilon \equiv \frac{1-t^2}{1+t^2}$. Within these assumptions and Eq.(21), we can find that the critical magnetic field ratio is the temperature-independent parameter as

$$\eta(t) = \eta(t=0) \qquad (23)$$

And we can get

$$B_{c2}(t) = B_{c2}(0)\frac{1-t^2}{1+t^2} . \qquad (24)$$

Here $B_{c2}(t)$ and $B_{c2}(0)$ are a temperature-dependence and zero-temperature upper critical field, respectively. This equation is the same formula of Ref.[19] that in one-band GL scenery.

Recently, Shanenko et al.[33] derived the extended GL- formalism for two-band superconductor that improved the validity of GL theory below $T_c$. They showed $\alpha \equiv (1-t) + \frac{1}{2}(1-t)^2$. To find $B_{c2}(T)$ that valid close to $T = 0$, we make the assumption that $\alpha \equiv p(1-t) + \frac{q}{2}(1-t)^2$ and $\varepsilon \equiv p(1-t)+\frac{q}{2}(1-t)^2$, where $p,q$ are the constant. The critical magnetic field ratio is also the temperature-independence parameter but the temperature-dependence upper critical magnetic field is

$$B_{c2}(t)=B_{c2}(0)\{p(1-t)+\frac{q}{2}(1-t)^2\} \ . \qquad (25)$$

## 3.Results and Discussions

In this paper, we are interested in the two-band magnetic superconductor found in Fe-base superconductors [11]. They show the very high upper critical magnetic field that the theory model is need for helping to predict the upper critical magnetic field. Zhu et al.[32] synthesized the iron-based superconductor $LaFeAsO_{0.9}F_{0.1-\delta}$ and found that the upper critical field $H_{c2}$ at zero-temperature was about 50 T by using one-band GL equation, $H_{c2}=\frac{\phi_0}{2\pi\xi^2}$,which $\phi_0$ is the flux quanta and $\xi$ is the coherence length, $\xi^2 \equiv \frac{1+t^2}{1-t^2}$. The lower critical of this material less than 20 Oe was found. Putti et al.[34] presented the overview of the superconducting main properties of Fe-based superconductors. The anisotropy with the huge upper critical field was discussed. The magnetoresistance measurements the upper critical field of polycrystalline La-1111($LaFeAs(O,F)$) up to 45 T [35] indicated a $\mu_0 H_{c2}(0)$ value larger than 60 T which corresponds to a small coherence length of order of a few nm and the London penetration depth of about 200 nm. The $H_{c2}$ shows the anomalous properties, similar to that observed in dirty $MgB_2$ [36,37] suggesting that superconductivity in Fe-based resulted from at least two bands. The $H_{c2}$'s extrapolation of Sm-1111($SmFeAs(O,F)$) yield $\mu_0 H_{c2}(T=0)\approx 400T$. The antiferromagnetism and the paramagnetism played an important role in these materials.

According to above data, we estimate the materials' properties as $\kappa_0 \approx 100$, $B_{c1} \approx 20$ Oe, $\chi \approx 0$ ,and $\chi' \neq 0$. The Eq.(20),Eq.(24) and Eq.(25) are used to investigate the critical field of Fe-base superconductors. Although the anisotropic function is included in our calculation that found in $\kappa_0$ and $\eta_0$ but it does not affect in our calculation, $\kappa_0$ and $\eta_0$ are the temperature-independence parameters.

Figure1.shows that higher value of the critical field ratio may be found in superconductor with diamagnetism($\chi<0$). According to the magnetic properties of Fe-base superconductors that show the antiferromagntism and paramagnetism that $\chi \approx 0$, we think that the differential susceptibility $(\chi')$ may affect to this property. So the critical magnetic ratio versus the differential susceptibility $(\chi')$ is also investigated and shown in Figure 2. We find that the higher value of critical magnetic ratio can be found in the negative differential susceptibility region $(\chi'<0)$. Then we can conclude that the very high upper critical field of Fe-base

superconductor( $\chi \approx 0$ ) can be found in the negative differential susceptibility region.

The temperature-dependent of upper critical field, Eq.(24) and Eq.(25), is shown in Figure 3. The curves are agreed with the experimental data that $\frac{B_{c2}(T=0)}{B_{c2}(0)} = 1$ and $\frac{B_{c2}(T=T_c)}{B_{c2}(0)} = 0$. The difference of Eq.(24) and Eq.(25) are in the region, $0 < T < T_c$. However, the Eq.(24) can be included in Eq.(25) depending on the parameter $p$ and $q$. The experimenter can use Eq.(24) and Eq.(25) to predict the zero-temperature upper critical field of the Fe-based superconductors and the other two-band superconductors.

## 4.Conclusions

The upper critical field, the lower critical field, the critical magnetic field ratio and the temperature-dependent upper critical field of an anisotropic two-band magnetic superconductor in GL scenery are derived analytically. Although there are many parameters in the two-band model, our results look simplifying. We find that the very high value of upper critical field of Fe-base superconductors can be found in the negative differential susceptibility region. The arbitrary constants are grouped by separating into temperature dependent and non-dependent part. The temperature-dependent upper critical field of Fe-based superconductors is investigated by using the temperature-dependent of $\alpha$ and $\varepsilon$ parameter in free energy. The temperature-dependent upper critical field is presented in two formulas as in the empirical view and in the GL two-band view that agrees with the experiment results.

## Acknowledgements

The authors would like to thank Professor Dr.Suthat Yoksan for the useful discussions. The financial support of the Srinakharinwirot University, Phathumwan Institute of Technology and ThEP center are acknowledged.

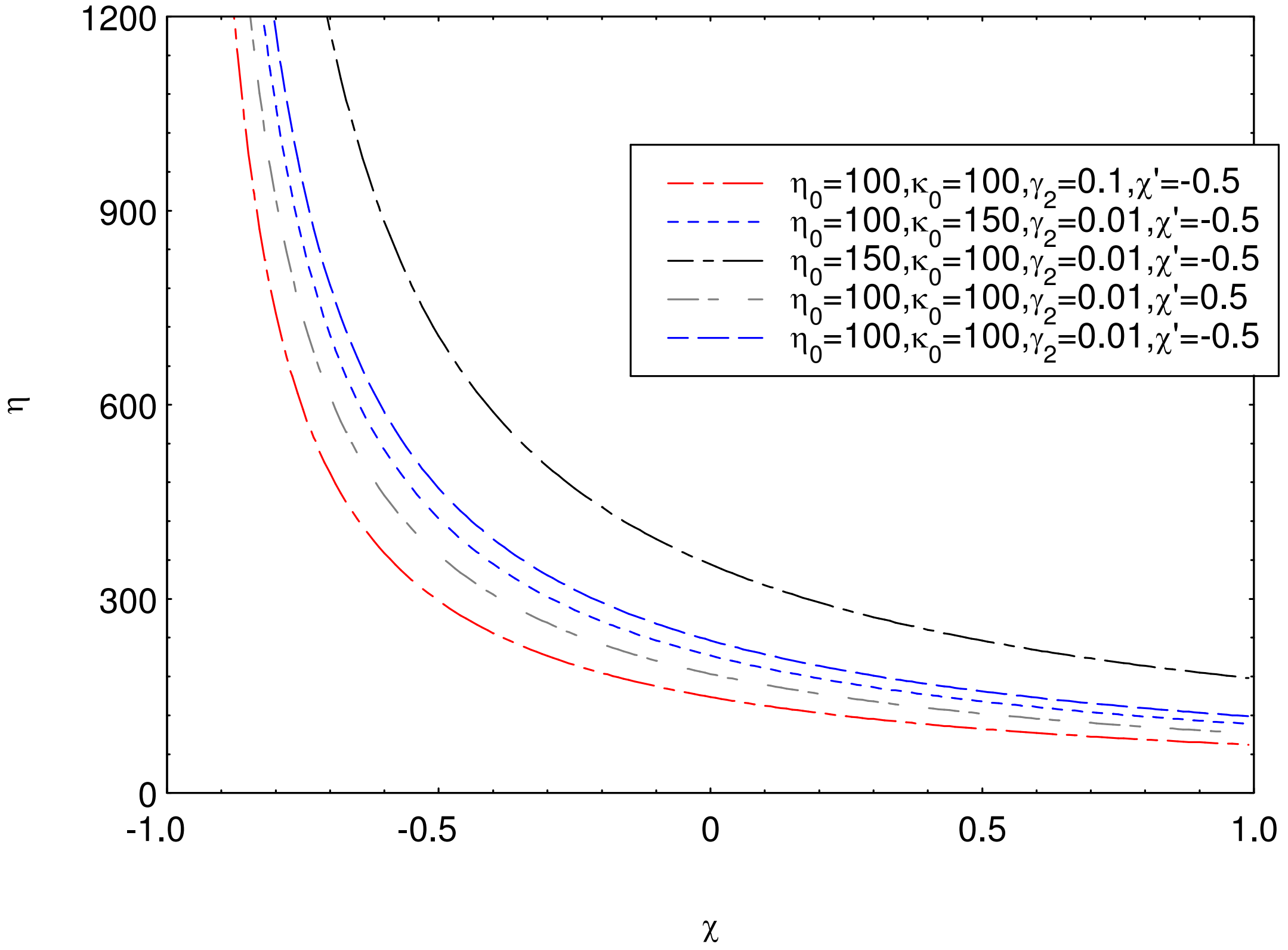

Figure 1. Shown the critical magnetic field ratio($\eta$)versus susceptibility$(\chi)$.

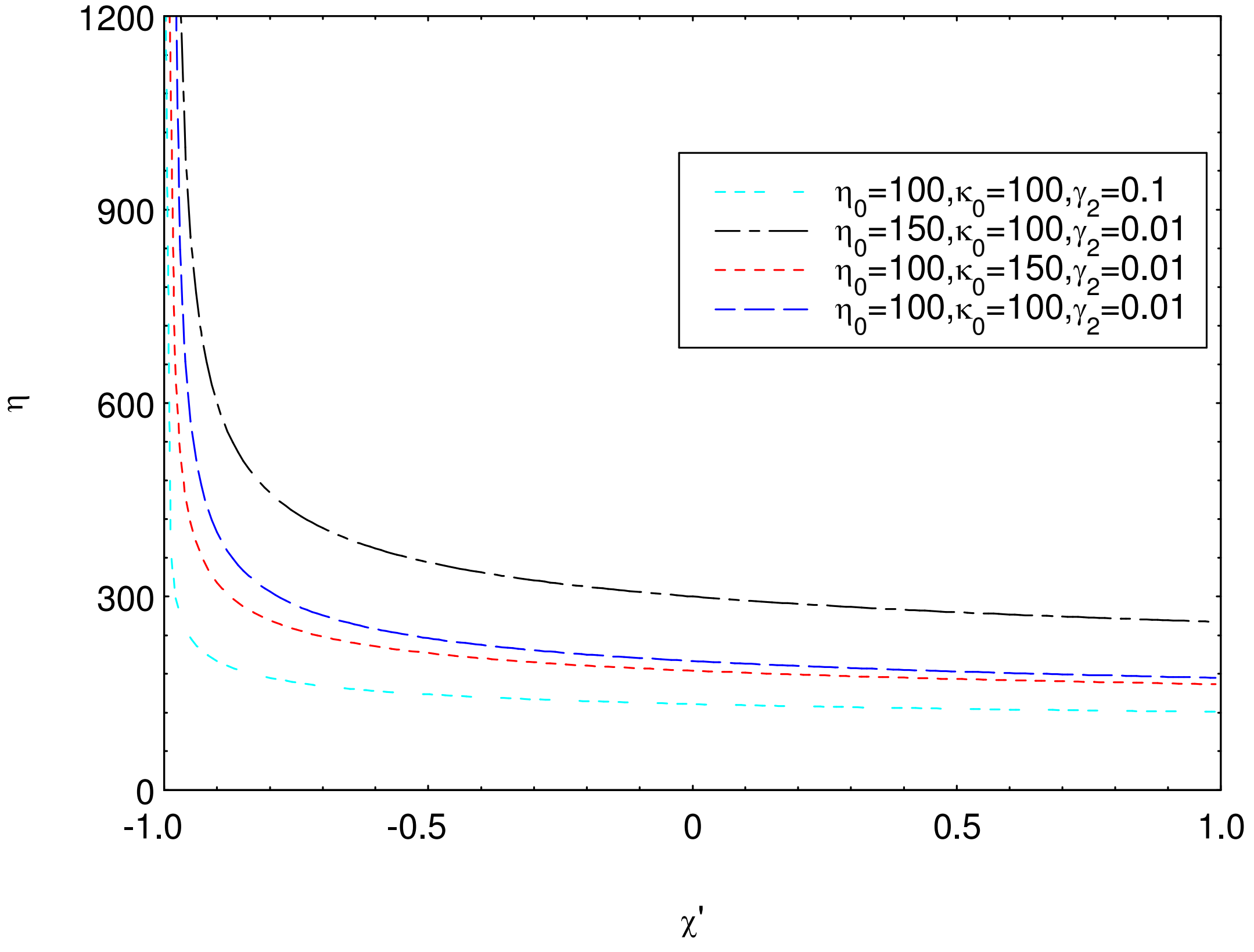

Figure 2. Shown the critical magnetic field ratio($\eta$) versus differential susceptibility $(\chi')$ with $\chi = 0$.

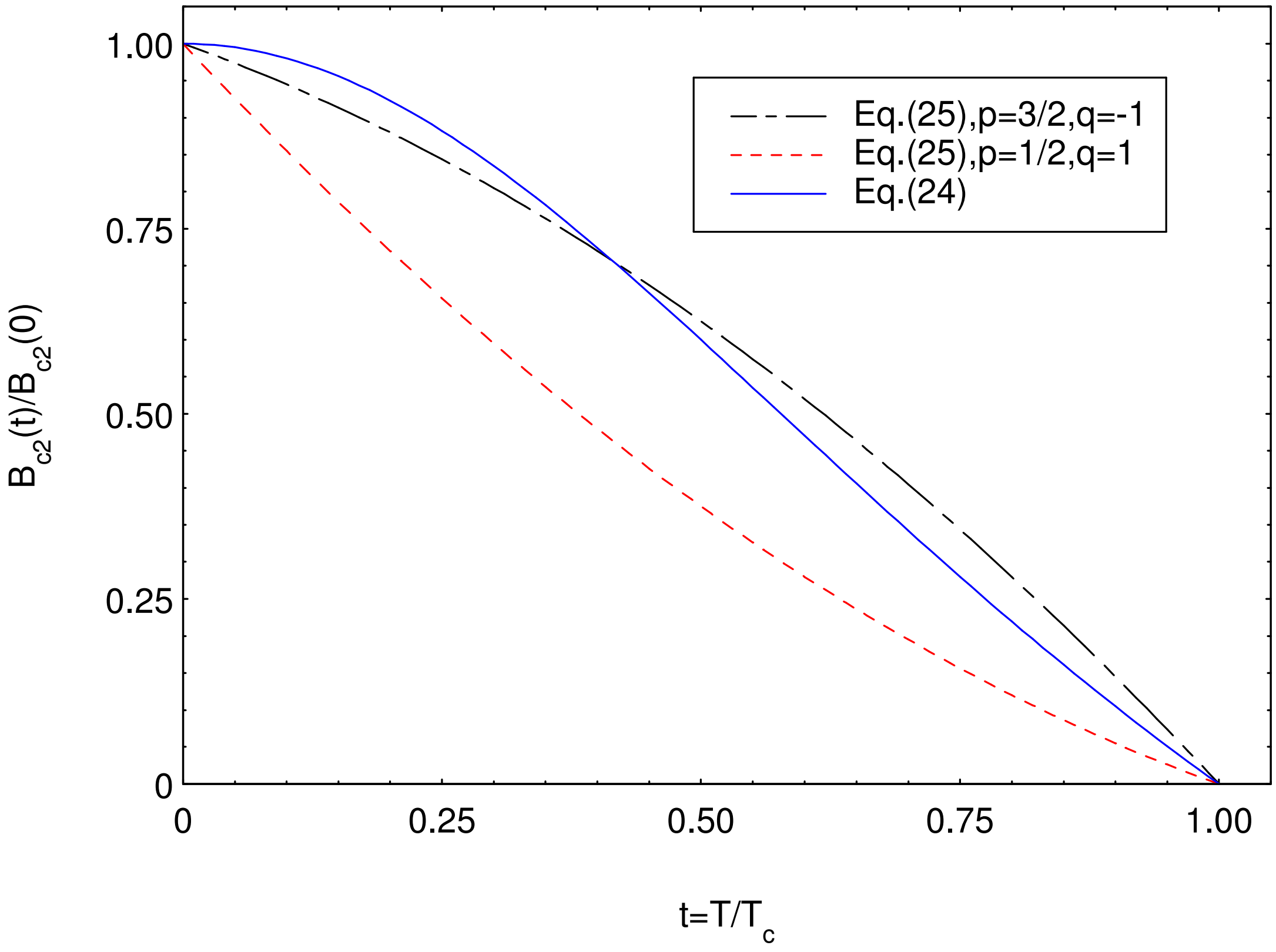

Figure 3. Shown the upper critical magnetic field versus temperature.